\documentclass{paper}
\usepackage{authblk}
\usepackage{hyperref}

\usepackage{graphicx}
\usepackage{amsmath}
\usepackage{slashed}

\newcommand{\bea}{\begin{eqnarray}}
\newcommand{\eea}{\end{eqnarray}}
\newcommand{\beq}{\begin{equation}}
{
\newcommand{\eeq}{\end{equation}}
\newcommand{\ec}{\end{center}}
\newcommand{\bc}{\begin{center}}

\newcommand{\pdir}{p\kern -5.2pt\raise 0.2ex\hbox {/}}

\newcommand{\vdir}{v\kern -5.75pt\raise 0.15ex\hbox {/}}
\newcommand{\kdir}{k\kern -5.75pt\raise 0.15ex\hbox {/}}
\newcommand{\epsdir}{\epsilon\kern -5.0pt\raise 0.15ex\hbox {/}}
\newcommand{\bvdir}{\bar{v}\kern -5.75pt\raise 0.15ex\hbox {/}}
\newcommand{\Ddir}{D\kern -7.75pt\raise 0.20ex\hbox {/}}
\newcommand{\Adir}{A\kern -7.75pt\raise 0.20ex\hbox {/}}
\newcommand{\ldir}{l\kern -5.0pt\raise 0.2ex\hbox{/}}
\newcommand{\varepsdir}{\varepsilon\kern -5.5pt\raise 0.15ex\hbox{/}}

\makeatother

\title{Leptoquarks in B-meson anomalies: simplified
models and HL-LHC discovery prospects}

\author{ Natascia Vignaroli}
 
 \affil{\small Dipartimento di Fisica ``E. Fermi", Universit\`{a} di Pisa, Italy}
 
\begin{document}

\maketitle

\begin{abstract}
 I will review simplified models with leptoquarks, which can explain recent anomalies in $B$-meson physics, and I will indicate the High-Luminosity LHC prospects for testing these theories, with a special focus to the efficient channel of pair leptoquark production in the $t\bar t$ plus missing energy final state.
\end{abstract}

%%%%%%%%%%%%%%%%%%%%%%%%%%%%%%%%%%%%%%%%%%%%%%%%%%
\section{Introduction}
\label{sec:intro}

A variety of theories beyond the Standard Model (BSM), as Pati-Salam model \cite{Pati:1974yy}, grand unification theories \cite{Georgi:1974sy} and BSM composite dynamics \cite{Pelaggi:2017wzr}, predict the existence of hypothetical particles carrying both lepton and baryon number, the so-called leptoquarks (LQs). These particles caught recently a special attention from the high energy physics community, since they represent the best candidates \cite{Fajfer:2012jt, Alonso:2015sja, Fajfer:2015ycq, Barbieri:2015yvd, Becirevic:2016yqi, Cai:2017wry, Aebischer:2019mlg, Angelescu:2018tyl, Calibbi:2017qbu, Alvarez:2018gxs}  to explain anomalies in flavor physics observed by  experiments on $B$-meson decays:  Belle \cite{Huschle:2015rga, Hirose:2016wfn, Sato:2016svk, Abdesselam:2016cgx}, Babar \cite{Lees:2012xj, Lees:2013uzd} and by LHCb \cite{Aaij:2015yra, Aaij:2014ora, 1705.05802}. In particular, the experiments find the indication of  lepton flavor universality violation in the ratio observables $R_{D^{(*)}}$, at about 4$\sigma$ level (by combining the results of the different experiments), and $R_{K^{(*)}}$. The most precise measurement of $R_{K^{(*)}}$ to date, by LHCb \cite{Aaij:2019wad}, shows a deviation of 2.5$\sigma$ from the Standard Model prediction. 
 It is really appealing that the anomalies can be explained simultaneously by models with LQs in the TeV range \cite{DiLuzio:2017chi}, thus in the reach of the LHC.  The optimization of the search strategies for LQs at the LHC is thus very important to enlighten the physics behind the flavor anomalies and in general for seeking BSM physics.\\

%%%%%%%%%%%%%%%%%%%%%%%%%%%%%%%%%%%%%%%%%%%%%%%%%%
\section{Simplified models for leptoquarks}
\label{sec:setup}

Motivated by the $B$-physics anomalies, we focus on two representative models: (i) the scalar LQ $S_3=(\mathbf{\bar{3}},\mathbf{3},1/3)$, where we indicate the SM quantum numbers, $(SU(3)_c,SU(2)_L,U(1)_Y)$, with the electric charge, $Q=Y+T_3$, and the (ii) vector LQ $U_1=(\mathbf{3},\mathbf{1},2/3)$, which we describe now in detail:

\begin{itemize}
\item[•] $\underline{S_3=(\mathbf{\bar{3}},\mathbf{3},1/3)}$: 

The $S_3$ LQ has been considered in models addressing the $B$-physics anomalies with two scalar LQs~\cite{Becirevic:2018afm,Marzocca:2018wcf}. The Yukawa Lagrangian of the simplified model for $S_3$ reads~\cite{Dorsner:2016wpm}
\begin{equation}
\label{eq:S3model}
\mathcal{L}_{S_3}  = y_L^{ij} \, \overline{Q^C_{i}} i \tau_2 ( \tau_k S^k_3) L_{j}+\mathrm{h.c.}\,,
\end{equation}
\noindent where $\tau_k$ ($k=1,2,3$) denote the Pauli matrices, $S_3^k$ are the LQ triplet component and $y_L$ is a generic Yukawa matrix. It is assumed that an appropriate symmetry forbids LQ couplings to diquarks, which are tightly constrained by experimental limits on the proton lifetime. If we recast the above expression in terms of charge eigenstates, we obtain:
\begin{align}
\begin{split}\label{eq:lag-S3}
\mathcal{L}_{S_3} = &- y_L^{ij} \, \overline{d^C_{L\,i}} \nu_{L\,j}\, S_3^{(1/3)}-\sqrt{2} \, y_L^{ij} \, \overline{d^C_{L\,i}} \ell_{L\,j}\, S_3^{(4/3)}\\[0.4em]
&+\sqrt{2}\,\left(V^\ast y_L\right)^{ij}\, \overline{u^C_{L\,i}} \nu_{L\,j}\, S_3^{(-2/3)}-\left(V^\ast y_L\right)^{ij} \overline{u^C_{L\,i}} \ell_{L\,j}\, S_3^{(1/3)}+\mathrm{h.c.}\,,
\end{split}
\end{align}
where $V$ is the CKM matrix. The superscript denotes the electric charge of the LQ states. 
Note that the model allows for the LQ interaction with muon and bottom and with  muon and strange, which can mediate a process accounting for the anomaly in $R_{K^{(*)}}$, and also an $S_3$ interaction with a top and a neutrino, which is relevant for the LQ search at colliders. This latter interaction leads to a $S_3 \to t\bar \nu$ decay with a branching fraction:
%%%%%%%%%%%%%%%%
\begin{equation}
\label{eq:br-s3}
\mathcal{B}(S_3^{(2/3)}\to t \bar{\nu}) \simeq \dfrac{(y_L \cdot y_L^\dagger)_{33}}{\displaystyle\sum_{i}\big{(}y_L \cdot y_L^\dagger \big{)}_{ii}}\, ,
\end{equation}
%%%%%%%%%%%%%%%%
where we adopted a compact notation, $(y_L \cdot y_L^\dagger)_{ii}\equiv \sum_j |y_{L}^{ij}|^2$. 

\item[•] $\underline{U_1=(\mathbf{3},\mathbf{1},2/3)}$: 

The $U_1$ model attracted a lot of attention because it can provide a simultaneous explanation to the anomalies in $b\to s$ and $b\to c$ transitions, with a single mediator~\cite{Buttazzo:2017ixm}. The most general Lagrangian consistent with the SM gauge symmetry allows couplings to both left-handed and right-handed fermions. If we neglect the interactions to right-handed fields, we have, in the mass eigenstate basis:

\begin{equation}
\mathcal{L}^{L}_{U_1}= \left(V^\ast x_L\right)^{ij}  \bar{u}_{L\,i} \gamma_\mu U_1^\mu \nu_{L\,j} + x^{ij}_L \bar{d}_{L\,i} \gamma_\mu U_1^\mu \ell_{L\,j} + \mathrm{h.c.} \ ,
\end{equation}

where $x_L^{ij}$ are Yukawa couplings, and we obtain that
%%%%%%%%%%%%%%%%
\begin{equation}
\mathcal{B}(U_1^{(2/3)}\to t \bar{\nu}) \simeq \mathcal{B}(U_1^{(2/3)}\to b \bar{\tau}) \simeq \frac{1}{2}\, \dfrac{(x_L \cdot x_L^\dagger)_{33}}{\displaystyle\sum_{i}\big{(}x_L \cdot x_L^\dagger \big{)}_{ii}}\,,
\end{equation}
%%%%%%%%%%%%%%%%
where we neglected fermion masses, similarly to Eq.~\eqref{eq:br-s3}.

The $U_1$ QCD interactions that control the $U_1$ pair production at colliders are determined by the kinetic terms:
\begin{equation}
\mathcal{L}^{kin}=-\frac{1}{2} U_1^{\dagger\mu\nu} U^{1}_{\mu\nu}- i \, g_s \, k \, U^{\dagger\mu}_{1} T^{a} U_{1}^{\nu} G^a_{\mu\nu} \, ,
\end{equation}
where $U^{\mu\nu}_1$ denotes the $U_1$ strength tensor and $k$ is a dimensionless parameter which depends on the ultraviolet completion of the model. We can identify the two scenarios of minimal coupling (MC), $k=0$, and the Yang-Mills (YM) case, $k=1$. 
\end{itemize} 

 In Table \ref{tab:lq-states}, we list different LQ states that can decay to $t\bar{\nu}$,  along with the corresponding operator, which can arise via interactions with a lepton doublet ($L$), or a right-handed neutrino ($\nu_R$). The third column of Table~\ref{tab:lq-states} indicates the maximal value of $\mathcal{B}(\mathrm{LQ}\to t\bar{\nu})$ allowed by gauge symmetry. 
In the following, we will assume that the dominant interactions are the ones to third-generation left-handed fermions, as suggested by the $B$-physics anomalies. In this case, the branching fractions to $t\nu$ will be $100\%$ for $S_3$ and $50\%$ for $U_1$, which are the most optimistic values.

%%%%%%%%%%%%%%%%%%%%%
\begin{table}[htbp!]
\renewcommand{\arraystretch}{1.7}
\centering
\begin{tabular}{|c|c|c|c|c|}\hline
Field  & Spin & Quantum Numbers &   Operators & $\mathcal{B}(\mathrm{LQ}\to t\bar{\nu})$ \\ \hline\hline
$R_2$  & $0$ & $(\mathbf{3},\mathbf{2},7/6)$ &  $\overline{u_R} R_2 i \tau_2 L$ & $\leq 0.5$ \\ 
$\widetilde{R_2}$  & $0$ & $(\mathbf{3},\mathbf{2},1/6)$ &  $\overline{Q} \widetilde{R_2} \nu_R$ & $\leq 1$ \\ 
$\bar{S}_1$  & $0$ & $(\overline{\mathbf{3}},\mathbf{1},-2/3)$ &  $\overline{u_R^C} \bar{S}_1 \nu_R$ & $\leq 1$ \\ 
$S_3$  & $0$ & $(\overline{\mathbf{3}},\mathbf{3},1/3)$ &  $\overline{Q^C} i \tau_2 \vec{\tau}\cdot \vec{S}_3L$ & $\leq 1$\\ \hline
$U_1$  & $1$ & $(\mathbf{3},\mathbf{1},2/3)$ &  $\overline{Q}\gamma_\mu U_1^\mu L\,,$~$\overline{u_R}\gamma_\mu U_1^\mu \nu_R$ & $\leq 0.5\,,~1$ \\ 
$\widetilde{V_2}$  & $1$ & $(\overline{\mathbf{3}},\mathbf{2},-1/6)$ & $\overline{u_R^C}\gamma_\mu \widetilde{V}^{\mu}_2 i \tau_2 L$\,,~$\overline{Q^C}\gamma_\mu i \tau_2 \widetilde{V}^\mu_2 \nu_R$  & $\leq 0.5\,,~1$ \\ 
$U_3$  & $1$ & $(\mathbf{3},\mathbf{3},2/3)$ & $\overline{Q}\gamma_\mu \vec{\tau}\cdot \vec{U}^\mu_3 L$  & $\leq 0.5$  \\ \hline
\end{tabular}
\caption{ \small Classification of the LQ states that can decay to $t\bar{\nu}$, in terms of the SM quantum numbers, $(SU(3)_c,SU(2)_L,Y)$, with $Q=Y+T_3$. We adopt the same notation of Ref.~\cite{Dorsner:2016wpm} and we omit color, weak isospin and flavor indices for simplicity. The last column corresponds to the maximal value of $\mathcal{B}(\mathrm{LQ}\to t \bar{\nu})$, as allowed by gauge symmetries. In the cases where interactions to lepton doublets ($L$) and right-handed neutrinos ($\nu_R$) are both allowed, i.e.~for the models $U_1$ and $\widetilde{V}_2$, we give the maximal branching fraction assuming only interactions to $L$ or $\nu_R$, respectively.}
\label{tab:lq-states} 
\end{table}

\section{LQ phenomenology at hadron colliders}

 The general LQ phenomenology at hadron colliders has been 
 explored in \cite{Mohapatra:1984aq} and more recently in \cite{Dorsner:2016wpm, Dorsner:2018ynv, Diaz:2017lit}. The relevant processes at the LHC are pair production of LQs driven by QCD interactions, single production mediated by model-dependent couplings of the LQs to leptons and quarks, $y\backslash x$, and the LQ exchange in the $t$-channel leading to high-$p_T$ dilepton final states, which depends quadratically on the couplings $y\backslash x$. Since the three processes depend differently on the $y\backslash x$ couplings, they can provide complementary probes at the LHC of different regions of the coupling-mass parameter space of the LQ models.
Several searches, which give bounds on the LQ masses, have been performed by ATLAS and CMS so far. The strongest limits on 2/3-charged third-generation LQs are currently set by the CMS analysis in \cite{Sirunyan:2019xwh}, which considered pair produced LQs each decaying to a neutrino and a top, bottom, or light-flavor quark and used 137 fb$^{-1}$ of data at a center of mass energy $\sqrt{s}=13$ TeV.  A vector LQ decaying 50\% to $t\nu$ is excluded by this analysis for masses below 1550 GeV, in the Yang-Mills (YM) case, and for masses below 1225 GeV in the minimal coupling (MC) scenario. A scalar LQ decaying 100\% to $t\nu$ is excluded up to masses of 1140 GeV. 
In the following, we summarize the main results of the analysis in \cite{Vignaroli:2018lpq}, which tried to improve the search strategy for LQs and estimated the sensitivity of the LHC at a collision energy of 14 TeV and at high luminosity. 
The study  in \cite{Vignaroli:2018lpq}  considers pair produced vector and scalar LQs each decaying into a top and a neutrino, leading to a final state of two tops plus missing energy. This channel, due to  a peculiar topology and to the possibility of exploiting the top tagging to disentangle the signal from the background, proves to be very powerful and it represents one of the best channels to probe LQs involved in the explanation of the flavor anomalies.

%%%%%%%%%%%%%%%%%%%%%%%%%%%%%%%%%%%%%%%%%%%%%%%%%%
\section{Search strategy in the $\bf t \bar{t}$ plus missing energy channel}
\label{sec:search}

We summarize in this section the main results of the study  in \cite{Vignaroli:2018lpq}, which outline a search strategy at the 14 TeV LHC for pair-produced scalar and vector LQs, decaying each into a top quark and a neutrino. In particular, the analysys considers the $U_1$ and $S_3$ LQs introduced in section \ref{sec:setup}, assuming a decay branching ratio into $t\nu$ of 50\% for $U_1$ and of $100\%$ for $S_3$. The final state is given by two tops decaying hadronically plus missing energy.  
The main background consists of $Z+\mathrm{jets}$ events where the $Z$ decays to neutrinos and leads to missing energy. Minor backgrounds come from $W+\mathrm{jets}$ and $t\bar t$ events, where a leptonic decaying $W$ leads to missing energy from the neutrino and a lost lepton \cite{Sirunyan:2018kzh}. 

Signal and background events are simulated at leading order with MadGraph5\_aMC@NLO \cite{Alwall:2014hca}. Events are then passed to Pythia \cite{Sjostrand:2006za} for showering and hadronization. A smearing to the jet momenta is also applied in order to mimic detector effects \cite{Ovyn:2009tx}. Signal events are generated via UFO files \cite{Degrande:2011ua}, created by using Feynrules \cite{Christensen:2008py}. For the case of the scalar LQ $S_3$, correction factors to the cross section values are applied, which account for QCD next-to-leading-order effects. They are calculate by using the code in \cite{Dorsner:2018ynv}. Jets are clustered with Fastjet \cite{Cacciari:2011ma} by using an anti-kt algorithm \cite{Cacciari:2008gp}. A large cone size, $R=1.0$, is chosen in order to optimize the top reconstruction procedure. 

The signal is characterized by large missing transverse energy, $\slashed{E}_T$, and at least two fat-jets, coming from the hadronic decays of the two tops. Considering these signal features, as a first step of the analysis, the events are accepted if they satisfy the conditions:
\begin{equation}
\slashed{E}_T > 250 \, \text{GeV}\,, \qquad n_j\geq2 \; \, (p_T \, j > 30 \,  \text{GeV} \, , \,  |\eta_j|<5)  \, ,\qquad \ \textsf{lep veto}\,,
\end{equation}
with $n_j$ denoting the number of jets satisfying the $p_T$ and rapidity requirements. Events are rejected if at least one isolated lepton, either a muon or an electron, with $p_T>$ 10 GeV and in the central region $|\eta|<$ 2.5 is found (\textsf{lep veto}). 

A crucial part of the search strategy in \cite{Vignaroli:2018lpq} relies on the reconstruction of both of the two tops in the final state. A simple reconstruction procedure is applied, which basically consists on cutting the fat-jets invariant mass around the top mass. Indeed, since the jets are clustered on a relatively large cone size and the tops in the signal are boosted, most of the top decay products are collected in a single fat-jet. Details are provided in \cite{Vignaroli:2018lpq}. The efficiency of the top pair tagging is of about 20\% for the signal, while the background is rejected by a factor of about 1.4$\cdot 10^3$. Only the events with two top tagged jets are then selected. Once having identified the two tops,  
several observables are constructed based on them, which can efficiently discriminate the LQ signals from the background. The selection is then completed by applying cuts on these ``top observables". One of these observables is inspired by the $M_{T2}$ variable commonly used by experimental searches \cite{Sirunyan:2017kqq}. In \cite{Vignaroli:2018lpq} it is constructed upon the tops, instead on jets, and it is defined as
\begin{align}\label{eq:MT2}
\begin{split}
& M_{T2} \equiv \text{max} \left\{ M_{T\, t(1)} , M_{T \, t(2)} \right\}\,, \\
&\\[-0.2cm]
M_{T\, t(i)} & = \sqrt{ 2   \slashed{E}_T  \,  p_T \, t(i)\,  \big( 1- \Delta\phi(\slashed{E}\, , t(i))/\pi \big) }\,, \qquad i=1,2\,,
\end{split}
\end{align}
where $p_T \, t(1,2)$ is the transverse momentum of the top $t (1,2)$ and $\Delta\phi(\slashed{E} \, , t(1,2))$ denotes the azimuthal angular separation between the missing energy vector and the top $t (1,2)$. Other ``top observables" used as signal-to-background discriminants are the invariant mass of the system made of the two tops, $M_{tt}$, and the transverse momenta of the tops.
The signal selection is thus refined by imposing the cuts:

\begin{equation}\label{eq:cut-main}
\slashed{E}_T > 500 \, \text{GeV} \quad M_{tt} > 800 \, \text{GeV}
\end{equation}

\noindent which exploits the large missing energy and the large invariant mass of the top pair system in the signal events, and the two set of cuts on the transverse momenta of the tops and on the $M_{T2}$ variable: 

\begin{align}\label{eq:ref-cuts}
\begin{split}
& \text{\it loose}: \qquad M_{T2}>800 \, \text{GeV} \quad p_T \, t(1) > 500  \, \text{GeV} \quad p_T \, t(2) > 300 \,  \text{GeV} \,, \\
& \text{\it tight}: \qquad M_{T2}>1100 \, \text{GeV} \quad p_T \, t(1) > 700  \, \text{GeV} \quad p_T \, t(2) > 500 \,  \text{GeV} \,,
\end{split}
\end{align}

\noindent where the {\it loose}  ({\it tight})  selection is applied to signals with masses up to (above) 1.4 TeV.

%%%%%%%%%%%%%%%%%%%%%%%%%%%%%%%%%%%%%%%%%%%%%%%%%%
\section{HL-LHC reach}
\label{sec:reach}

\begin{figure}[t]
\centering
\includegraphics[width=0.5\linewidth]{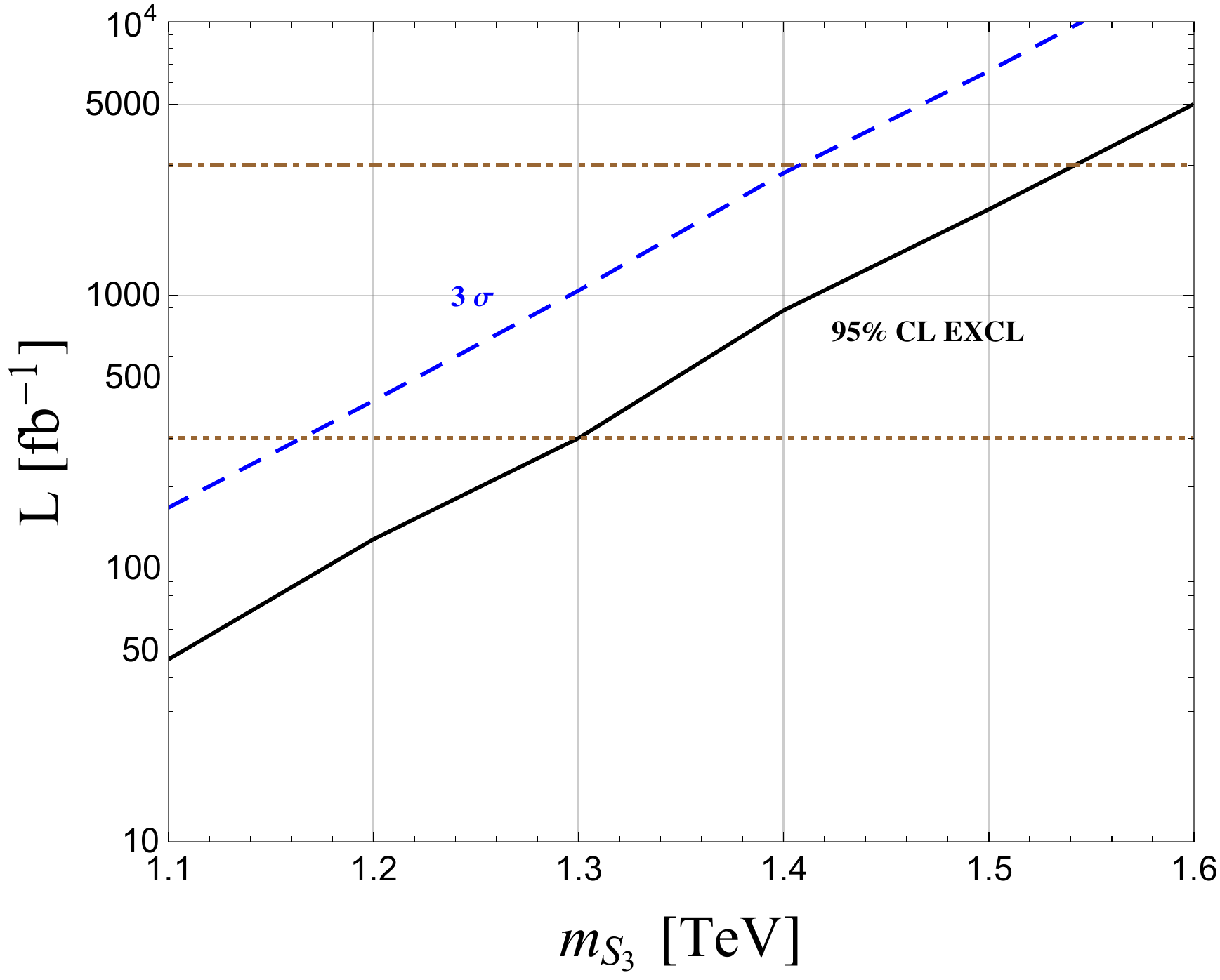} \\ [0.2cm]
\includegraphics[width=0.5\linewidth]{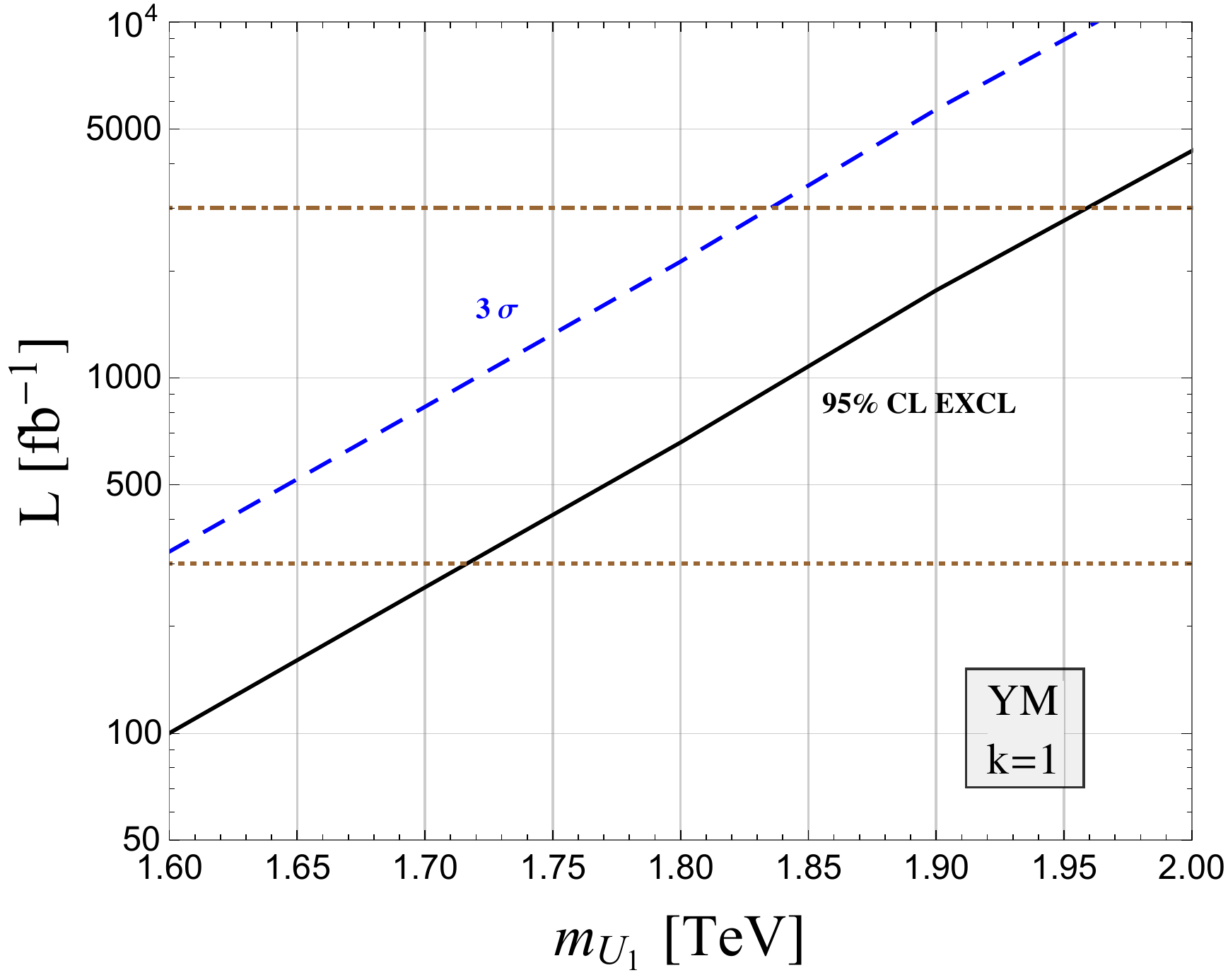}~\includegraphics[width=0.5\linewidth]{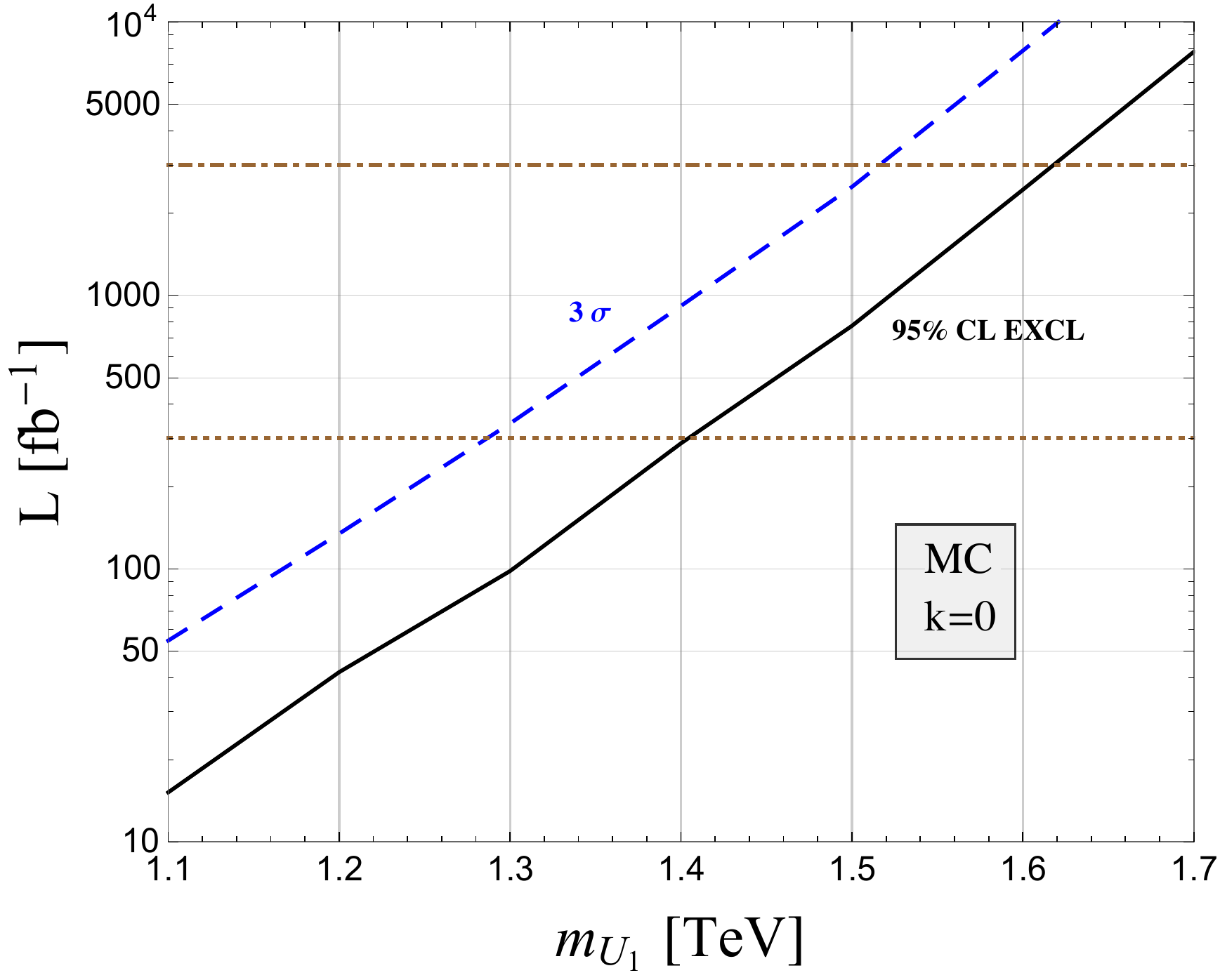} 
\caption{\small HL-LHC reach. Integrated luminosity required to exclude at 95\% C.L. (black line) or to observe at 3$\sigma$ (blue dashed line) a scalar LQ $S_3$ (upper plot) and a vector LQ $U_1$ (lower plots) as a function of their mass. For $U_1$, the plot on the left (right) refers to the YM (MC) scenario with $k=1 (0)$.}
\label{fig:reach}
\end{figure}

Fig. \ref{fig:reach}, taken from \cite{Vignaroli:2018lpq}, indicates the HL-LHC reach on vector and scalar LQs derived from the analysis summarized in the previous section. The results in \cite{Vignaroli:2018lpq} show that with 3 ab$^{-1}$ (300 fb$^{-1}$) the HL-LHC can exclude a vector LQ $U_1$ up to 1.96 TeV (1.72 TeV) or observe at 3$\sigma$ the corresponding signal for masses up to 1.83 TeV (1.6 TeV) in the YM case. In the MC scenario, $U_1$ LQs  up to 1.62 TeV (1.4 TeV) can be excluded with 3 ab$^{-1}$ (300 fb$^{-1}$). For the scalar LQ $S_3$, the exclusion reach extends up to 1.54 TeV (1.3 TeV) with 3 ab$^{-1}$ (300 fb$^{-1}$), while $S_3$ as heavy as 1.41 TeV (1.16 TeV) can be observed at 3$\sigma$. The study in \cite{Vignaroli:2018lpq} thus shows that the identification of the tops in the final state and the use of ``top observable" for the signal-to-background discrimination is very efficient to improve the LHC sensitivity to LQs.
Furthermore, the study applies a simple cut-and-count analysis so that we expect that these results are conservative. A more refined top reconstruction, making use for example of substructure techniques as ``jettiness" \cite{Thaler:2010tr, Stewart:2010tn} or a statistical analysis of the shape of the relevant distributions considered in \cite{Vignaroli:2018lpq} could augment the reach of the HL-LHC. 

%%%%%%%%%%%%%%%%%%%%%%%%%%%%%%%%%%%%%%%%%%%%%%%%%%

%%%%%%%%%%%%%%%%%%%%%%%%%%%%%%%%%%%%%%%%%%%%%%%%%%
\section{Conclusions}
\label{sec:conclusions}

LQs are interesting particles to be searched for at colliders. They are predicted in appealing BSM models and they represent the best candidates to accomodate $B$-physics anomalies.
The  $t\bar t$  plus missing energy channel from pair production of third-generation LQs proves to be one of the most efficient to discover LQs.
A dedicated search in the channel at the LHC, relying on the $t\bar t$ tagging, can significantly extend the reach. In particular, 
``top observables" constructed upon the tagged tops are useful to both discriminate the signal from the background and to characterize the signal.
The HL-LHC reach is wide on the parameter space of interesting models and in particular on the LQ models that can explain the flavor anomalies, as shown in \cite{Vignaroli:2018lpq}.

\end{document}